\documentclass[draft]{agujournal2019}
\usepackage{url} 
\usepackage[inline]{trackchanges} 
\usepackage{soul}
\usepackage{xcolor}

\usepackage{verbatim}

\draftfalse

\journalname{JGR: Space Physics}

\begin{document}

%
%


\title{The East-West Asymmetry of Particle Intensity in Energetic Storm Particle Events}

%
%




\authors{Zheyi Ding\affil{1},
          Gang Li\affil{2},
          Adolfo Santa Fe Dueñas\affil{3,4}, 
          Robert W. Ebert\affil{3,4},
          Nicolas Wijsen\affil{5,6},
          Stefaan Poedts \affil{1,7}
          }

\affiliation{1}{Centre for mathematical Plasma Astrophysics, KU Leuven, 3001 Leuven, Belgium}
\affiliation{2}{Department of Space Science and CSPAR, University of Alabama in Huntsville, Huntsville, AL 35899, USA}
\affiliation{3}{Southwest Research Institute, San Antonio, TX, USA }
\affiliation{4}{Department of Physics and Astronomy, University of Texas at San Antonio, San Antonio, TX, USA}
\affiliation{5}{NASA, Goddard Space Flight Center, Heliophysics Science Division, Greenbelt, MD 20771, USA}
\affiliation{6}{Department of Astronomy, University of Maryland, College Park, MD 20742, USA}
\affiliation{7}{Institute of Physics, University of Maria Curie-Sk{\l}odowska, Pl.\ M.\ Curie-Sk{\l}odowska 5, 20-031 Lublin, Poland}





\correspondingauthor{Gang Li}{gangli.uahuntsville@gmail.com}




\begin{keypoints}
\item East-West Asymmetry of particle intensity is often found in ESP events. We propose continuous shock acceleration can lead to this asymmetry.
\item Continuous acceleration depends on shock geometry and through this, the injection efficiency plays a central role for this asymmetry
\item We simulate this asymmetry using the iPATH model and compared our simulation results with observations
\end{keypoints}

%
%

%
%


\begin{abstract}
We examine the East-West asymmetry of the peak intensity in energetic storm particle (ESP) events using the improved Particle Acceleration and Transport in the Heliosphere (iPATH) model. We find that injection efficiency peaks east of the nose of coronal mass ejection shock where the shock exhibits a quasi-parallel geometry. We show that the peak intensity at the eastern flank is generally larger than that at the western flank and it positively correlates with the injection efficiency. We also examine this asymmetry for heavy ions, which depends sensitively on the ion energy. Comparison between the modelling results with the measurements of ESP events at 1 au shows a reasonable agreement. We suggest that the injection efficiency can be  a primary factor leading to the East-West asymmetry of the peak intensity in ESP events. Additionally, the charge-to-mass ($Q/A$) dependence of the maximum particle energy affects this asymmetry for heavy ions.
\end{abstract}

\section*{Plain Language Summary}

Energetic storm particle (ESP) events occur when coronal mass ejection-driven shocks pass a spacecraft, leading to abrupt increases in particle intensity and posing severe radiation hazards to astronauts and spacecraft. These enhancements are usually interpreted as the result of a local particle acceleration process. Therefore, in-situ measurements of ESP events provide a great opportunity to investigate the shock acceleration mechanism.  Recent observations from multiple spacecraft show an East-West asymmetry in the peak intensity of ESP events, with significantly different intensities observed on the eastern and the western shock flank. In this study, we use the 2D improved Particle Acceleration and Transport in the Heliosphere (iPATH) model to investigate the East-West asymmetry of particle intensity in ESP events. We find that the injection efficiency, which depends on the shock geometry, is the key parameter responsible for this East-West asymmetry.

%
%
\section{Introduction} \label{sec:intro}
Solar energetic particles (SEP) accelerated by the coronal mass ejection (CME) shocks are known as gradual SEP events. During gradual events, the enhancements of particle intensity associated with the passage of the interplanetary shock near spacecraft are referred to as energetic storm particle (ESP) events \cite{Bryant1962}. The properties of ESP events observed at 1 au, and their correlations with shock properties and upstream conditions have been widely studied. For instance, \citeA{Desai2003} found significant correlations between the interplanetary shock abundances and the ambient superthermal ions. \citeA{Reames2012ApJ...757...93R} suggested that the shock speed correlates best with the particle intensities in ESP events. Additionally, \citeA{ebert2016ApJ...831..153E} examined seven multiple-spacecraft ESP events and found that the peak intensities near the shock nose are larger than at the flank of the shock. More recently, \citeA{Duenas2022} showed that heavy ion peak intensities and spectra at 1 au are organized by longitude relative to their source flare location, which appeared to have an East-West asymmetric distribution of the peak intensity. This asymmetry refers to the difference of particle intensities between the eastern and the western shock flank. Unlike the East-West asymmetry of particle intensity in large SEP events, which is affected by the extended shock acceleration and the transport effects \cite<e.g.,>{Lario2006, Strauss2017,Xie2019,ding2022modelling},  ESP events are typically interpreted as a direct consequence of a local shock acceleration process. Thus, the East-West asymmetry of the particle intensity in ESP events may provide important insights into the underlying acceleration mechanism.

In ESP events, the diffusive shock acceleration (DSA; \citeA{Axford1977ICRC...11..132A}) mechanism is regarded as a primary mechanism for accelerating protons and ions at the shock. In the DSA theory,  particles are accelerated by moving in the turbulent magnetic fields near the shock and traversing the shock many times to gain energy. A controversial issue in DSA is how particles are injected at the shock from thermal or superthermal plasma, known as the ``injection problem" \cite<See e.g., >{Desai+2016}. For particles to participate in the DSA process, their speeds must exceed an injection threshold so that they can scatter diffusively across the shock, which refers to the injection speed $V_{\rm inj}$. A classical threshold of $V_{\rm inj}$ in DSA is the de Hoffmann–Teller speed \cite{le2009time}: 
\begin{equation}
V_{\rm inj} = U_{\rm up}/\cos\theta_{\rm BN}, 
\label{eq:HT injection spd}
\end{equation}
where $U_{\rm up}$ is the upstream flow speed in the shock frame and $\theta_{\rm BN}$ is the angle between the upstream magnetic field and the shock normal direction. The physical meaning of $V_{\rm inj}$ is clear: an ion moving along the upstream magnetic field with a speed $v>V_{\rm inj}$ can stay in front of the shock, thus participating in the shock acceleration process.  However, considering the gyrophase degree of freedom, the injection speed at a quasi-parallel shock should be larger than that derived from Equation~(\ref{eq:HT injection spd}).
An alternative approach by \citeA{giacalone+1999} (also used in \citeA{Zank+2006}) is to require the particle anisotropy to be small when using the Parker transport equation (see more discussion in Section~\ref{sec:model}). Both forms explicitly show the injection speed as a function of shock obliquity. With the knowledge of injection energy, the injection efficiency is defined as the ratio of integral number density above the injection energy to upstream flow density. Therefore, it also depends on the shock obliquity. If the injection energy increases with increasing shock obliquity angle, then the injection efficiency decreases. This feature has been addressed in \citeA{ellison1995acceleration,Li+2012,Battarbee2013A&A...558A.110B}.      
 The injection efficiency of the seed population is not only important for determining particle intensity but also plays a  crucial role in particle acceleration. The intensity of the self-generated waves, which govern the maximum particle energy attainable at a shock, is proportional to the number of injected particles \cite{Zank+etal+2000,Rice+2003,Li+2003}. Therefore, the injection efficiency not only determines the injected particle number density but also affects the maximum particle energy. \citeA{Li+2012} examined shock obliquity dependence of injection efficiency and its correlation with maximum particle energy. They suggested that a quasi-parallel shock is more efficient in accelerating particles. At 1 au, CME-driven shocks typically have a quasi-parallel geometry at the eastern flank and a quasi-perpendicular geometry at the western flank under nominal Parker magnetic field conditions. Consequently, particle intensity and maximum particle energy at the western and eastern flanks can differ. Due to the limited number of spacecraft at different longitudes at 1 au,  it is often impossible to observe an individual CME-driven shock and its ESP phase by multiple spacecraft ($>3$) that are well longitudinally separated. In contrast, numerical models provide an effective approach to simulate the longitudinal variation of particle intensity in ESP events at a CME-driven shock.

In this study, we investigate the East-West asymmetry of the peak intensity in ESP events using the two-dimensional (2D) improved Particle Acceleration and Transport in the Heliosphere (iPATH) model \cite{Hu+etal+2017}. In the original one-dimensional (1D) PATH model \cite{Zank+etal+2000,Rice+2003,Li+2003,Li+2005a}, which only considered the propagation of the CME-driven shock through a uniform solar wind only in the radial direction and adopted the steady-state DSA solution at different times. The iPATH model is extended from the PATH model, which includes the evolution of the shock obliquity in the ecliptic plane. Thus, it is capable of simulating the time intensity profiles and spectra at multiple spacecraft simultaneously.   
Later \citeA{Ding2020} and \citeA{Li2021} have examined two ground level enhancement (GLE) events during solar cycle 24 using the iPATH model and showed reasonable agreements with observations at multiple spacecraft. Recently, \citeA{ding2022modelling} utilized the iPATH model to examine the East-West asymmetry of particle fluence in large SEP events. They suggested that this asymmetry is a result of the effects of the extended shock acceleration and the geometry of the magnetic field. These works have demonstrated the capability of iPATH in including the necessary physical processes of particle acceleration at the shock and the propagation of energetic particles.

In the following, the injection of the seed population and the DSA solution in the iPATH model are described in Section~\ref{sec:model}. The results of the model and observation are shown in Section~\ref{sec:results}. A conclusion is given in Section~\ref{sec:conclusion}.

\section{Model} \label{sec:model} 

One important parameter in understanding the SEP events is the cut-off energy for particle acceleration at the shock. In the steady-state solution of DSA mechanism \cite{Drury+1983,Zank+etal+2000}, the shock parameters do not change significantly during the shock dynamic time scale $t_{\rm dyn}$, which, following \citeA{Li2017ScChD} equals,
\begin{equation}
t_{\rm dyn}=\rm min \left\{\frac{R}{dR/dt},\frac{B}{dB/dt},\frac{N}{dN/dt} \right\},
\label{eq:tdyn}
\end{equation}
where $R$, $B$ and $N$ are the radial distance from the sun, the magnetic field magnitude and the particle number density at the shock downstream, respectively.
The maximum particle energy is obtained by balancing the acceleration time with  $t_{\rm dyn}$,
\begin{equation}
t_{\rm dyn} = \int_{p_{\rm inj}}^{p_{\rm max}}\frac{3s}{s-1}\frac{\kappa}{U_{\rm up}^{2}}\frac{1}{p}dp,
\label{eq:dynamic time}
\end{equation}
where $p$ is the particle momentum, $p_{\rm inj}$ and $p_{\rm max}$ are the injection momentum and the maximum particle momentum, $s$ is the shock compression ratio, $\kappa$ is the particle diffusion coefficient, $U_{\rm up}$ is the upstream solar wind speed in the shock frame. In the DSA theory it is not considered how the particles are injected into the shock. \citeA{giacalone+1999,Zank+2006} calculated the injection speed by requiring the total anisotropy $\xi$ to be smaller than $1$: 
\begin{equation}
\xi=3\frac{|\mathbf{F}|}{vf}  = 3 \frac{ U_{\rm up}}{v}\left[(\frac{\beta}{3}-1)^2 + \frac{\kappa_{\rm Bohm}^2 \sin^2 \theta_{\rm BN} + (\kappa_{\parallel}-\kappa_{\perp})^2 \sin^2 \theta_{\rm BN} \cos^2 \theta_{\rm BN}}
{(\kappa_{\parallel} \cos^2 \theta_{\rm BN} + \kappa_{\perp} \sin^2 \theta_{\rm BN})^2}\right]^{1/2},
\label{eq:anisotropy}
\end{equation}
where $\mathbf{F}=  -\mathbf{\kappa} \cdot \nabla f - { U}_{\rm up}\frac{p}{3}  \frac{\partial f} {\partial p}$ is the streaming flux in the shock frame and $f$ is the particle distribution function. The second term is due to the Compton-Getting effect.  $v$ is the particle speed and  $\beta=3s/(s-1)$,  $\kappa_{\parallel}$ and $\kappa_{\perp}$ are parallel and perpendicular diffusion coefficient, and $\kappa_{\rm Bohm} = vr_{L}/3$ is the Bohm diffusion coefficient ($r_{L}$ is the gyroradius). See \citeA{Zank+2006} for a complete derivation of the particle anisotropy. Comparing the equation in \citeA{giacalone+1999}, we note that the dependence of $\beta$ in Equation~(\ref{eq:anisotropy}) is a result of the Compton-Getting effect in the shock frame. However, this approach needs to know the diffusion coefficients $\kappa$, but the injection speed is required to calculate $\kappa$ in the iPATH model when considering the amplified wave intensity at the shock front. To avoid this dilemma, if we assume that $\xi=1$ and $\kappa_{\perp},\kappa_{\rm Bohm} \ll \kappa_{\parallel}$, the injection speed is approximated by 
 \begin{equation}
 V_{\rm inj} =   U_{\rm up} \left[(\beta-3)+ 3\tan \theta_{\rm BN} \right].
\label{eq:injection speed_zank}
 \end{equation} 
In the above approximation, we use $a+b$ to approximate $\sqrt{a^2+b^2}$, where $a=\beta/3-1$ and $b=\tan \theta_{\rm BN}$. This approximation works well for both cases of $a\gg b$ and $a \ll b$, and overestimates $\sqrt{a^2+b^2}$ by a factor of $1.4$ when $a=b$. This simplification does not affect the correlation between injection efficiency and shock parameters, nor does it change the primary conclusion of our study.  It is worth noting that the assumption that $\kappa_{\perp}$ and $\kappa_{\rm Bohm}$ are much smaller than $\kappa_{\parallel}$ only holds for small magnetic fluctuations. The simplest model for $\kappa_{\parallel}$ is Bohm diffusion, which assumes that the parallel mean free path cannot be smaller than the Larmor radius of the charged particles. Generally speaking, for small turbulent fluctuations, $\kappa_{\parallel}$ is much larger than $\kappa_{\rm Bohm}$. Regarding $\kappa_{\perp}$, the perpendicular transport of a particle is primarily governed by  the field line random walk around the mean magnetic field $B$. If the magnetic fluctuations are small,  it is often assumed that $\kappa_{\perp}$ is much smaller than $\kappa_{\parallel}$. However, if magnetic fluctuations are strong  ($\delta B \sim B$), the isotropic spatial diffusion leads to $\kappa_{\perp}$ being approximately equal to $\kappa_{\parallel}$, and  $\kappa_{\parallel}$  approaches $\kappa_{\rm Bohm}$, so Equation~(\ref{eq:injection speed_zank}) is not valid anymore.
For a strong shock ($\beta=4$), Equation~(\ref{eq:injection speed_zank}) reduces to Equation~(\ref{eq:HT injection spd}) when $\theta_{\rm BN} =0^{\circ}$, but Equation~(\ref{eq:injection speed_zank}) is about three times larger than  Equation~(\ref{eq:HT injection spd}) when  $\theta_{\rm BN} \to 90^{\circ}$. This can be circumvented if we replace the $3\tan \theta_{\rm BN}$ by 
$\tan \theta_{\rm BN}$ in Equation~(\ref{eq:injection speed_zank}). This replacement does not affect the injection at a parallel shock, but for oblique shocks, it yields a smaller value than using Equation~(\ref{eq:injection speed_zank}).
This can be improved by using the following ansatz: 
 \begin{equation}
 V_{\rm inj} =   U_{\rm up} \left[(\beta-3 \eta )+ \tan \theta_{\rm BN} \right],
\label{eq:injection speed_eta}
 \end{equation} 
where $\eta$ is a parameter to regulate the injection speed. 
From the discussion below equation~(\ref{eq:HT injection spd}), we see that a proper choice of 
$\eta$ should be smaller than 1. Taking $\eta=1$ and $\beta=4$, Equation~(\ref{eq:injection speed_eta}) yields a similar behavior to Equation~(\ref{eq:HT injection spd})  as $\theta_{\rm BN} \rightarrow 0^{\circ}$ and $ \theta_{\rm BN}  \rightarrow 90^{\circ}$. By considering different values of $\eta$, we can estimate the various thresholds of injection speed at the parallel shock. We note that the choice of injection speed/energy is associated with the resulting maximum particle energy since injection energy affects injection efficiency and therefore the amplified wave intensity. A comparison of the injection energy and maximum energy for the aforementioned injection forms  can be found in \ref{appendix:injection}. 
Choosing $\eta=1/3$, we recover the injection speed used in \citeA{Li+2012,Hu+etal+2017}. 
We assume that this equation is valid for $0^{\circ}<\theta_{\rm BN}<85^{\circ}$.  At nearly perpendicular shocks ($\theta_{\rm BN} > 85^{\circ}$), cross-field diffusion via field line random walk plays a major role in particle injection. 
Some observational and numerical studies suggest a lower threshold of injection speed at quasi-perpendicular shocks due to field line meandering allowing particles to cross the shock front multiple times \cite<e.g.,>{Giacalone2005ApJIrregular,NeergaardParker2014ApJ...782...52N}. 
\citeA{Giacalone2005ApJthermal} have shown that thermal population can be accelerated by perpendicular shocks with sufficient pre-existing large-scale turbulence. To address particle injection at nearly perpendicular shocks, we must consider the perpendicular diffusion coefficient in the injection form. Thus Equation~(\ref{eq:injection speed_eta}) is not applicable at perpendicular shocks and we do not consider the cases of $\theta_{\rm BN} > 85^{\circ}$ in this work. We note that in reality, due to the stochastic nature of the IMF \cite{Bian+2021, BianLi2022}, having a portion of a shock with $\theta_{\rm BN} > 85^{\circ}$ for an extended period of time can be very rare.

Equation~(\ref{eq:injection speed_eta}) shows that the injection speed increases from a quasi-parallel shock to a quasi-perpendicular shock and from a higher shock compression ratio to a lower one. We note that the injection speed also depends on the shock speed, since $U_{\rm up}$ is the upstream speed in the shock frame. The total injected particle number density is proportional to the shock speed associated with the volume it swept during a time step, but it is not a key point in explaining the East-West asymmetry of the peak intensity. Particles above the injection speed are regarded as the seed particles, which are assumed to be a single power law distribution,
 \begin{equation}
f\left(E\right) = f\left(E_{\rm inj}^{0}\right) \left(\frac{E}{E_{\rm inj}^{0}} \right)^{-\delta},
 \end{equation}
 where $E$ is the particle energy, and $E_{\rm inj}^{0}$ is the injection energy at the strongest parallel shock (i.e., $\theta_{\rm BN}=0^{\circ}$ and $s=4$), $\delta$ is the spectral index. For different events $\delta$ generally varies from 1.0 to 3.5 in the ambient solar wind prior to the arrival of the shocks \cite{Desai2004,Desai2006SSRv..124..261D}.   In this work, we use $\delta=3.5$ as the highest limit, but it is a free parameter that can be constrained by ambient populations.  The spectral index is an important parameter to adjust the magnitude of injection  efficiency and asymmetry of injection efficiency as discussed in \ref{appendix:injection}.  The ratio of injection particle number density at an oblique part of the shock ($N_{1}$) and at the parallel part of the shock ($N_{0}$) is given by,
 \begin{equation}
 \frac{N_{1}}{N_{0}} \propto\left(\frac{E_{\rm inj}^{1}}{E_{\rm inj}^{0}} \right)^{1-\delta},
 \label{eq:density ratio}
 \end{equation}  
 where $E_{\rm inj}^{1}$ is the injection energy at an oblique shock. Early observational studies suggested 0.5\% $\sim$ 1\% thermal solar wind protons are accelerated at interplanetary shocks \cite{Gosling1981JGR....86..547G,Gloeckler1994JGR....9917637G}. Recent iPATH modelling works for realistic SEP events also suggested the injection efficiency can vary from several 0.1\% to 1\% \cite{Ding2020,Li2021,ding2022A&A...668A..71D}.  Following  the previous works \cite{Zank+etal+2000,Li+2012,Hu+etal+2017}, the injection efficiency $\epsilon_{0}$ at a strong parallel shock (i.e., $\beta=4$ and $\theta_{\rm BN}=0$) is set to be $1\%$ in this work.  Using Equations~(\ref{eq:injection speed_eta}) and (\ref{eq:density ratio}) and $\eta=1/3$, the injection efficiency $\epsilon_{1}$ at an oblique shock becomes,
 \begin{equation}
 \epsilon_{1} = \epsilon_{0} \left(\frac{(\beta-1)+\tan\theta_{\rm BN}}{3}\right)^{2-2\delta}.
  \label{eq:injection_efficy}
 \end{equation}
Equation~(\ref{eq:injection_efficy}) shows that the injection efficiency depends on the shock obliquity, the shock compression ratio and the spectral index of the seed population. This is the key point in explaining the East-West asymmetry of the peak intensity of ESP events, we discuss it more in Section \ref{sec:results}. 

In \citeA{Ding2020}, the instantaneous particle distribution function $f$ at the shock front $\mathbf{r}$ at time step $t_k$ is described by a power-law with an exponential tail, 
\begin{equation}
f(\mathbf{r}, p,t_k) = c_1\epsilon_{\mathbf{r}}n_{\mathbf{r}}p^{-\beta}H[p-p_{\rm inj, \mathbf{r}}] 
\exp\left(-\frac{E}{E_{b,\mathbf{r}}}\right),
\label{eq:fp}
\end{equation}
where $\epsilon_{\bf{r}}$ is the injection efficiency, $n_{\bf{r}}$ is the upstream solar wind density,
$p_{\rm inj,\bf{r}}$ is the particle injection momentum, and $E_{b,\mathbf{r}}$ is the kinetic energy that corresponds to a maximum proton momentum $p_{\rm max,\bf{r}}$. $H$ is the Heaviside function.  $c_1$ is a normalization constant,
\begin{equation}
c_1 = 1/\int_{p_{\rm inj,\bf{r}}}^{+\infty}p^{-\beta} H[p-p_{\rm inj,\bf{r}}]\exp\left(-\frac{E}{E_{b,\bf{r}}}\right)d^3p.
\end{equation}
Downstream of the shock, the accelerated particles advect and diffuse in the shell model of iPATH. The detailed description of the shell model can be found in \citeA{Zank+etal+2000,Hu+etal+2017}. Tracking particles in each shell allows one to compute the particle spectrum downstream of the shock as a function of time, including the ESP phase when the shock passes over the spacecraft. Some examples of ESP events in the iPATH model can be found in \citeA{Fu+2019}. In this work, we focus on the peak intensity of ESP events and investigate the causes for the East-West asymmetry of the peak intensity in ESP events.

\section{Results} \label{sec:results} 
\begin{figure}[ht]
    \centering
    \includegraphics[width=\textwidth]{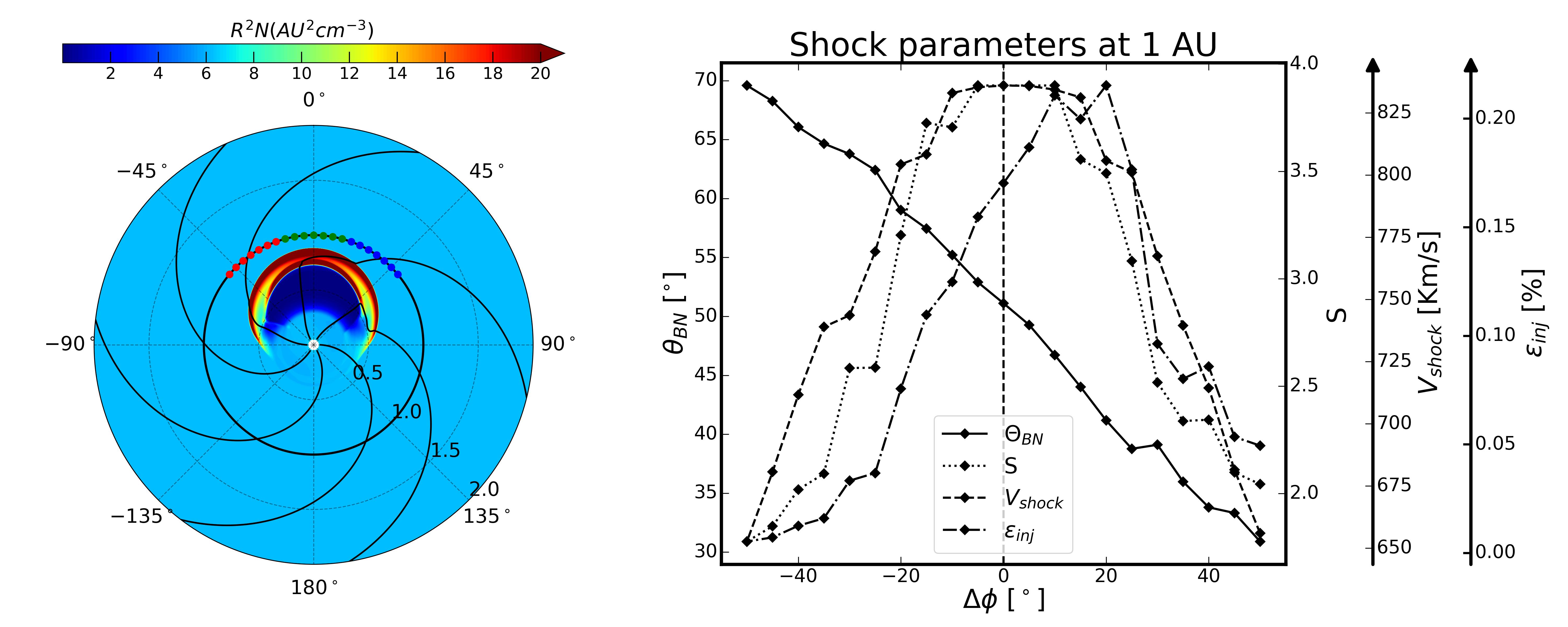}
    \caption{Left: the equatorial snapshot of the modelled scaled number density. 21 virtual observers are divided into 3 groups marked by red, green and blue dots. Black curves represent Parker magnetic field lines. See text for details. Right: shock parameters at 1 au as a function of SC-CME deflection angle ($\Delta \phi$). The shock obliquity ($\theta_{\rm BN}$), the shock compression ratio ($s$), the shock speed ($V_{\rm shock}$) and the injection efficiency ($\epsilon_{\rm inj}$) are  denoted by solid, dotted, dashed and dash-dotted lines, respectively. To guide the eye, a dashed vertical line at the center $0^{\circ}$ is shown. }
    \label{fig:cme-shock}
 \end{figure}

By way of example, we consider a CME with an eruption speed  of $1600$ km/s and a width of $120$ degrees, launched at a heliocentric distance of $0.05$ au. The CME propagates into a uniform solar wind with a speed of $400$ km/s. The left panel of Figure~\ref{fig:cme-shock} shows the equatorial snapshot of the scaled number density $n r^2$ from 0.05 au to 2.0 au. The black curves represent nominal Parker field lines. The center of CME propagates towards $0^{\circ}$. A total of $21$ virtual observers at 1 au, separated by $5^{\circ}$ in longitude, are denoted by dots. In this work, instead of using the SC-flare angle to classify events as in the work of \citeA{Duenas2022}, we define an SC-CME deflection angle,  $\Delta \phi$, 
which is the longitudinal difference between the CME center and the spacecraft. Note that in observations, it is harder to decide the CME center direction than the flare location. If we assume the CME center and the flare have the same longitude, our definition of $\Delta \phi$ becomes that in \citeA{Duenas2022}. A positive (negative) $\Delta \phi$ represents that the spacecraft is located at the eastern (western) side of the CME center. To describe the shock geometry, we distribute the $21$ observers into three groups: the green dots, which correspond to the shock nose, have $\Delta \phi$ between $-15^{\circ}$ and $15^{\circ}$; the blue dots, which correspond to the eastern flank, have $\Delta \phi$ between $20^{\circ}$ and $50^{\circ}$; and the red dots, which correspond to the western flank, have $\Delta \phi$ between $-20^{\circ}$ and $-50^{\circ}$.
 
The right panel in Figure~\ref{fig:cme-shock} shows the shock parameters as a function of SC-CME deflection angles. The distributions of $s$ and $V_{\rm shock}$ are almost symmetrical with respect to $0^{\circ}$, but $\theta_{\rm BN}$ decreases from $70^{\circ}$ to $30^{\circ}$ as $\Delta \phi$ increases from $-50^{\circ}$ to $50^{\circ}$. The injection efficiency is calculated from these shock parameters using Equation~(\ref{eq:injection_efficy}). If the shock compression ratio and the spectral index of the seed population are the same, the injection efficiency decreases as the shock obliquity angle increases. Therefore, the distribution of injection efficiency is asymmetric and peaks around $\Delta \phi = 20^{\circ}$. This asymmetry is due to the longitudinal distribution of shock obliquity, indicating that the quasi-parallel shock is more suitable for seed particle injection.

 \begin{figure}[ht]
    \centering
    \includegraphics[width=\textwidth]{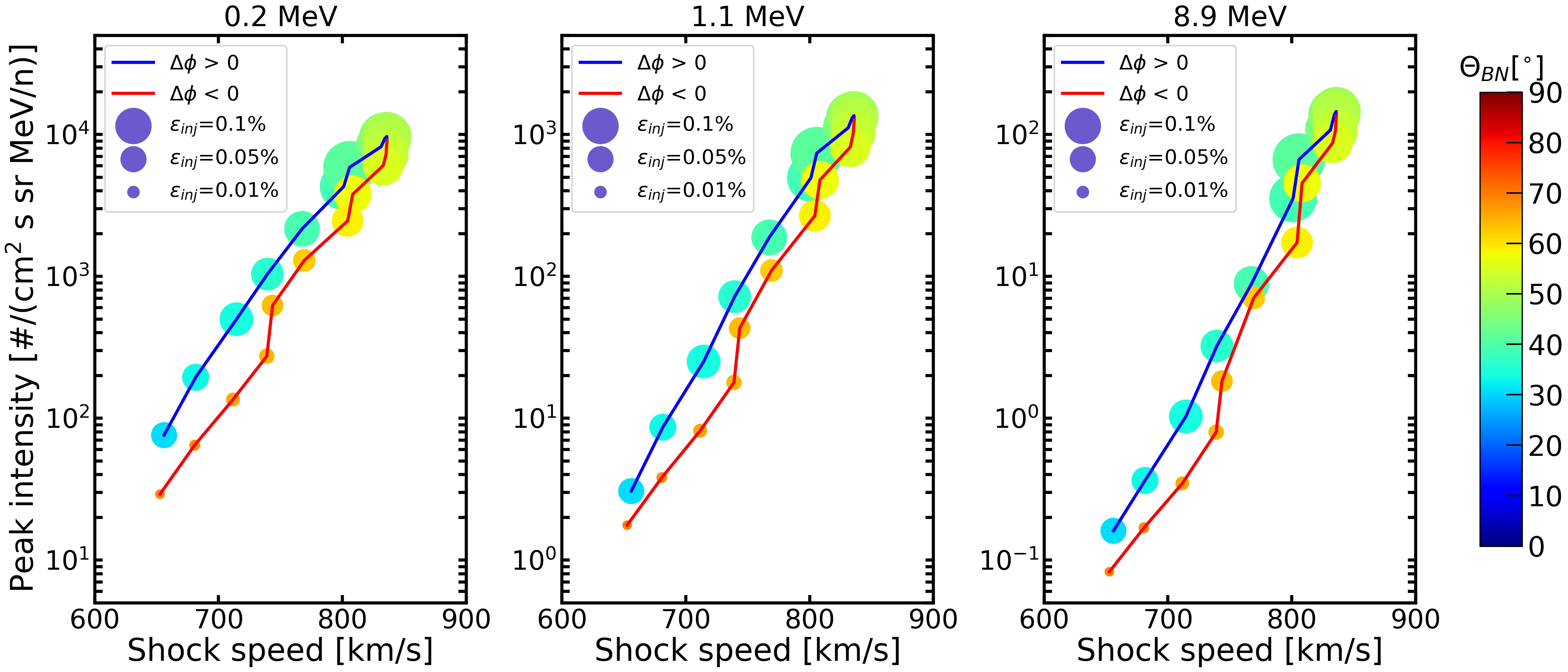}
    \caption{Peak proton intensity at $0.2$ MeV/n, $1.1$ MeV/n, $8.9$ MeV/n versus the shock speed recorded at each virtual observer at 1 au. 
    The shock obliquity and the injection efficiency are indicated by the color and the size of the dots. Their numerical values are labelled in Figure S1 in the Supporting Information. Blue (red) lines are for $\Delta \phi>0^{\circ}$ ($\Delta \phi<0^{\circ}$).  To guide the eye,  the size for the symbols corresponding to $\epsilon_{\rm inj}=0.1\%$, $\epsilon_{\rm inj}=0.05\%$ and $\epsilon_{\rm inj}=0.01\%$  are shown to the upper left. } 
    
    \label{fig:peakintensity-spd}
 \end{figure}
 
Figure~\ref{fig:peakintensity-spd} shows the proton peak intensity of ESP events at 21 virtual observers versus the shock speed for three energy channels of $0.2$ MeV/n, $1.1$ MeV/n and $8.9$ MeV/n. Since particle intensity is largely affected by the shock speed \cite{Reames2012ApJ...757...93R}, it is clearer to show the East-West asymmetry of the peak intensity as a function of the shock speed rather than  SC-CME deflection angles. The blue and red lines correspond to $\Delta \phi$ larger than $0^{\circ}$ and smaller than $0^{\circ}$. The colors of the dots show the shock obliquity angle, and the size of the dots indicates the injection efficiency. The larger dot represents the higher injection efficiency. It is clear that the peak intensity at the eastern flank ($\Delta \phi> 0^{\circ}$) is generally larger than that at the western flank ($\Delta \phi< 0^{\circ}$) for the shown energies. If we ignore the magnitude of peak intensity in three energy channels, the fluctuations of data points between different energies are similar. This is because the injection efficiency is energy-independent. In this case, between the eastern and the western flanks, the difference of peak intensity at a similar shock speed is around several times.  It is difficult to validate this difference from current observations due to the limited multiple-spacecraft events. In recent multiple-spacecraft studies of ESP events by \citeA{ebert2016ApJ...831..153E,Duenas2022}, for an individual event, the difference of peak intensity of $\sim 0.5$ MeV/n ions at two spacecraft is mostly smaller than a factor of $10$ and the shock speeds are different between two spacecraft. Therefore, we suppose that the magnitude difference of peak intensity may be small between East-West flanks with similar shock speeds. We note that the slope is larger in the channel of $8.9$ MeV/n since shock with low shock speed and low compression ratio is less efficient for accelerating particles to high energies.  

  \begin{figure}[ht]
    \centering
    \includegraphics[width=\textwidth]{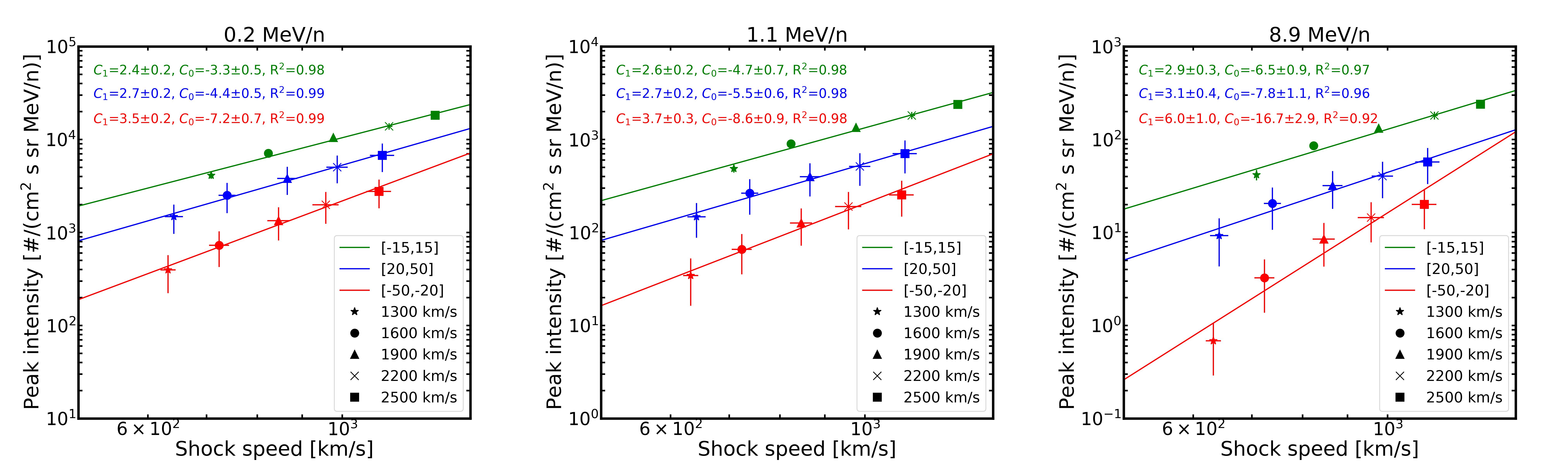}
    \caption{The average proton peak intensity at $0.2$ MeV/n, $1.1$ MeV/n, $8.9$ MeV/n versus the average shock speed recorded at 1 au. Five cases with different CME eruption speeds are labelled by different symbols. We note that the labels of different symbols represent the CME eruption speed.  The horizontal error bar is the standard error of the shock speed, and the vertical error bar is for the peak intensity. Three groups of virtual observers are denoted by green, blue and red, respectively. The fitting parameters $C_1$ and $C_0$ and the coefficient of determination $R^2$ are shown in the top left of each panel.     }
    \label{fig:protons}
 \end{figure}

To further investigate the East-West asymmetry of ESP events and to better compare to observational results, we consider five cases with different CME eruption speeds, which are $1300$ km/s, $1600$ km/s, $1900$ km/s, $2200$ km/s, and $2500$ km/s respectively. In the work of \citeA{Gopalswamy2010}, the average speeds  of CMEs with shocks are about 999 km/s and the fraction of partial/full halo CMEs (width $\ge 120$ degrees) is 88\%.  Of these CMEs, they found that the average speeds of CMEs with radio-loud shocks are 1237 km/s and the fraction of partial/full halo CMEs (width $\ge 120$ degrees) is 96\%. These radio-loud shocks are presumably the source of large SEP events. Furthermore, in the work of \citeA{Duenas2022}, the authors have found that that the angular width of ESP events has a clear threshold at near-sun CME speed to be $1300$ km/s. Above this speed, the ESP events show a significant longitudinal dependence. 
Therefore, we also choose the CME eruption speed larger than $1300$ km/s in our simulation, which is necessary to examine the ESP events in a wide longitudinal extent.  We further assume the width of CME to be a constant of $120^{\circ}$ in all cases to get adequate ESP events at the eastern and the western flanks of the shock. In general, the CME width does not intensively change the shock properties between the eastern and the western flanks. Consequently, we focus on the relationship between CME speeds and the peak intensity of ESP events.
 We divide 21 virtual observers into three groups as shown in Figure~\ref{fig:cme-shock}. We then average each group of observers' peak intensities and shock speeds. We choose three energy channels of $0.2$ MeV/n, $1.1$ MeV/n and $8.9$ MeV/n. The average peak intensities for five cases versus average shock speeds are shown in Figure~\ref{fig:protons}. In each panel, green, blue and red symbols represent the observers related to the shock nose, the eastern flank and the western flank. The labels of different symbols represent the different CME eruption speeds. This result allows us to compare the East-West asymmetry of the peak intensity to statistical results in observations. We utilize the same expression in \citeA{Duenas2022} to fit the relation between the shock speed ($V$) and the peak proton intensities ($I$): 
 \begin{equation}
\log_{10}(I) = C_{1} \log_{10}(V) + C_{0},
\label{eq:loglogfit}
 \end{equation}
where $C_{1}$ is the slope and $C_{0}$ is the y-intercept of the $log-log$ fit. The fitting parameters and the coefficient of determination $R^2$ are labelled in the upper left of each panel. For all three energies, the peak intensity is highest at shock nose ($\Delta \phi \in [-15^{\circ},15^{\circ}]$), and is always higher at the eastern flank ($\Delta \phi \in [20^{\circ},50^{\circ}]$) than at the western flank ($\Delta \phi \in [-50^{\circ},-20^{\circ}]$). We note that the highest peak intensity observed at central events is due to the high compression ratio ($>3$) near the shock nose in this study. As discussed in Section~\ref{sec:model}, a high compression ratio leads to a high injection efficiency. Between eastern and western events, the shock compression ratio is similar as shown in Figure~\ref{fig:cme-shock}, therefore the main difference in shock properties is the shock obliquity. The slope $C_{1}$ at the western shock flank is the largest, which suggests that the injection efficiency of the western shock flank is weaker than that at the shock nose or eastern flank. Furthermore, the slope difference between the eastern and western flanks becomes larger in higher energies. It suggests that quasi-parallel geometry at the eastern flank can accelerate protons to higher energies than  quasi-perpendicular geometry at the western flank.

  \begin{figure}[ht]
    \centering
    \includegraphics[width=\textwidth]{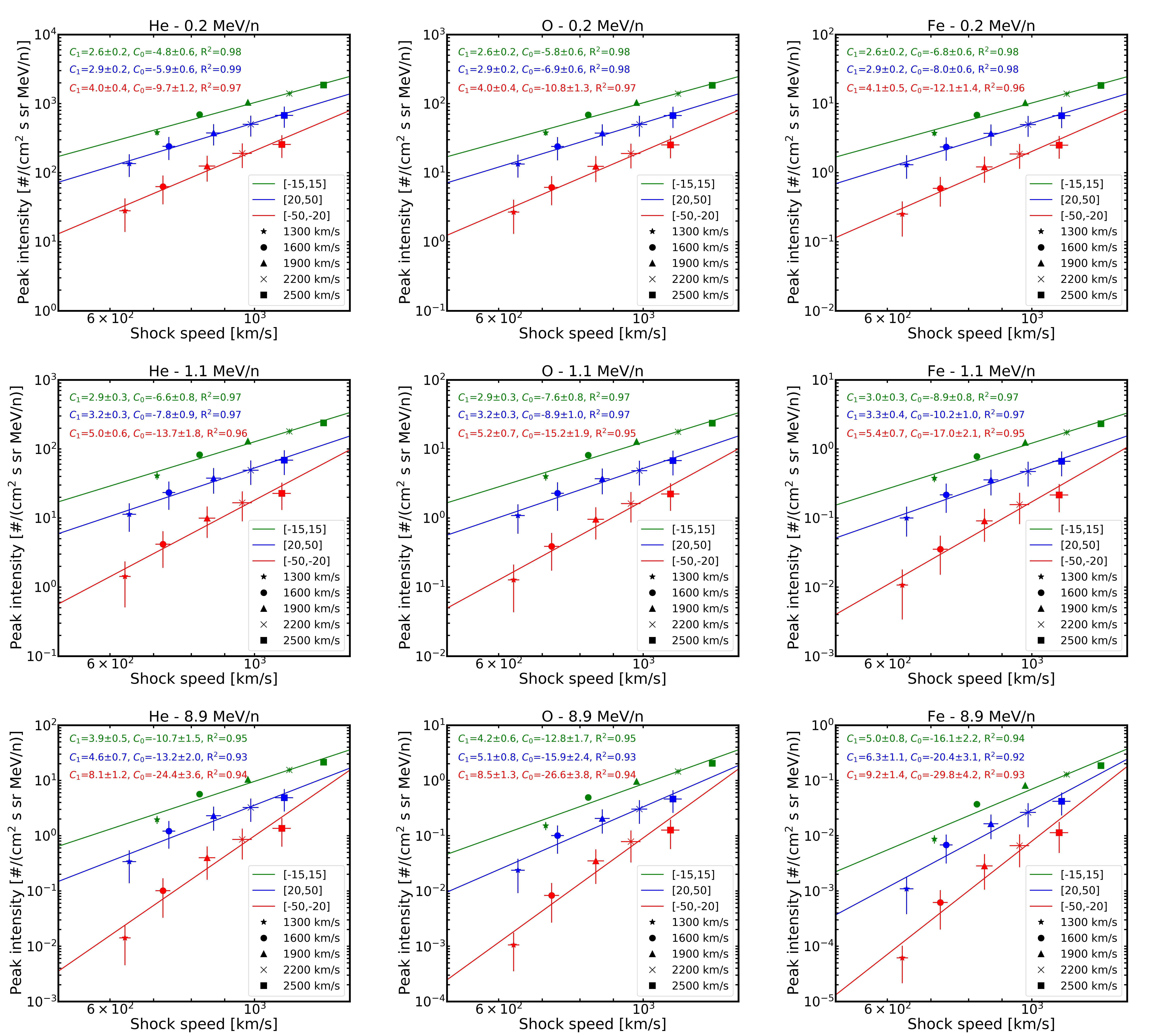}
    \caption{Same format as in Figure~\ref{fig:protons} but for He, O and Fe. The upper line is for $0.2$ MeV/n ions, the middle line is for $1.1$ MeV/n ions and the lower line is for $8.9$ MeV/n ions. We note that the labels of different symbols represent the CME eruption
speed.}
    \label{fig:ions}
 \end{figure}

We now examine the East-West asymmetry for heavy ions. 
At the shock front, ions resonate with the ambient or proton-excited waves. With the same pitch angle and same energy/nucleon, ions with a lower charge-to-mass ($Q/A$) ratio resonate with a smaller wave number, with $Q$ being the ion charge number and $A$ being the ion mass in the unit of proton mass. The iPATH model has considered the heavy ion acceleration through interacting with the wave turbulence generated by the streaming protons \cite{Li+2005a,li+2009}. \citeA{Li+2005a} showed that the maximum particle energy/nucleon could be related to the rigidity dependence of the diffusion coefficient and suggested a cut-off energy/nucleon  $\sim (Q/A)^2$ at a parallel shock.  Figure~\ref{fig:ions} shows the peak intensity at $0.$2 MeV/n, $1.1$ MeV/n and $8.9$ MeV/n for Helium (Fe), Oxygen (O) and Iron (Fe) as a function of shock speeds. The format of this figure is the same as Figure~\ref{fig:protons}. We also use Equation~(\ref{eq:loglogfit}) to fit the peak intensity for the three groups of observers. We can see that the East-West asymmetry of the peak intensity exists in three species of ions. At the energy of $0.2$ MeV/n and $1.1$ MeV/n, the slope $C_{1}$ for He, O and Fe are almost the same for all three groups of observers. This is because ions can be well accelerated to $1.1$ MeV/n in all cases. However, the slope $C_{1}$ shows large variations at the energy of $8.9$ MeV/n. For instance, for the eastern group (blue), $C_{1}= 4.6 \pm 0.7$ for He; $C_{1}=5.1 \pm 0.8$ for O; and $C_{1}=6.3 \pm 1.1$ for Fe. The typical value of $Q/A$ for He, O, and Fe in gradual SEP events are $2/4$, $7/16$ and $12/56$ respectively \cite{Klecker1999ICRC....6...83K,Desai2016ions}.  As the $Q/A$ value of ions decreases, their maximum energy also decreases, resulting in a steeper slope of the fitting line for ions with smaller $Q/A$.

  \begin{figure}[ht]
    \centering
    \includegraphics[width=\textwidth]{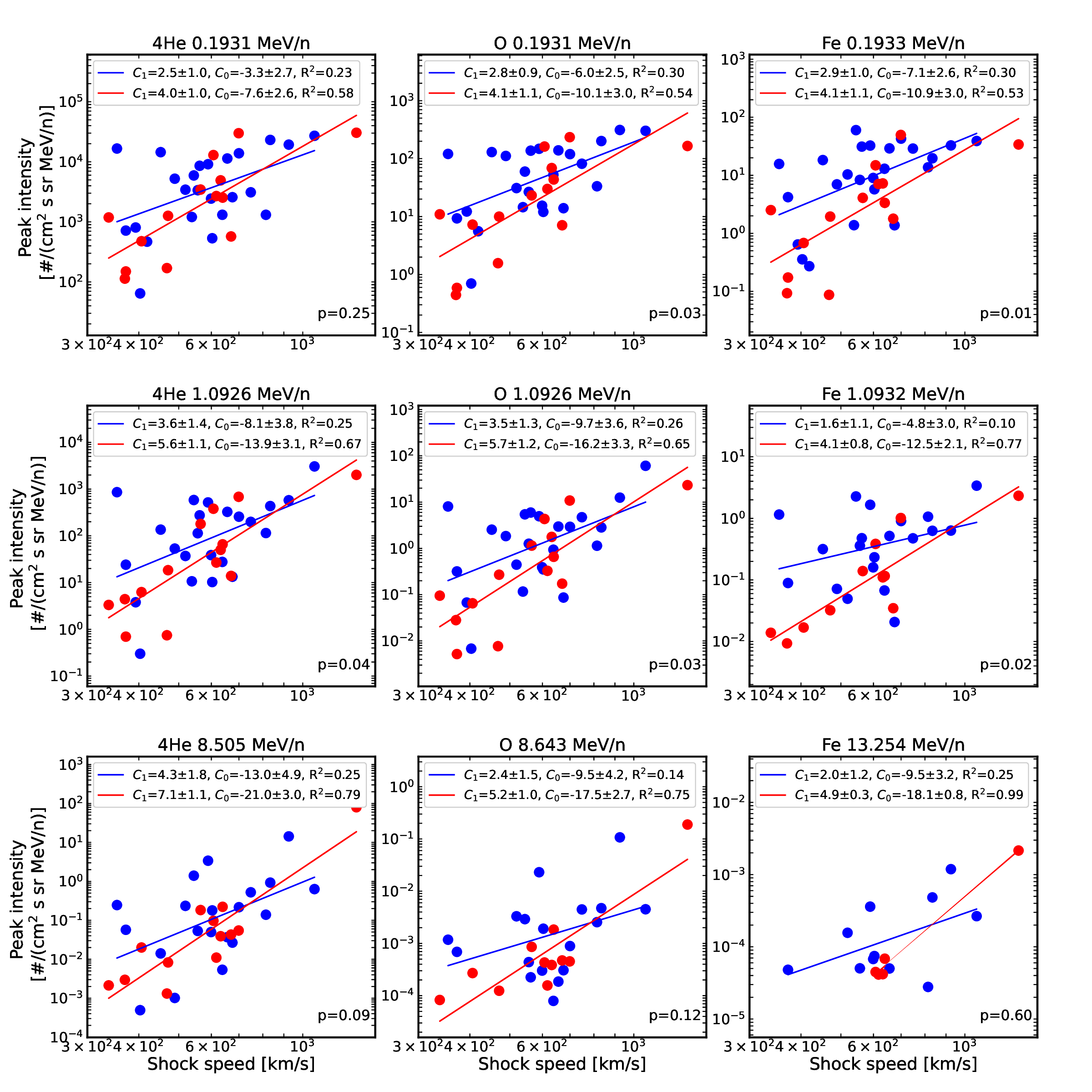}
    \caption{Peak intensity versus the IP shock speed at $0.2$ MeV/n (upper row), $1.1$ MeV/n (middle row) and $8.5$ MeV/n (lower row) for He, O and Fe in the observations. Western (eastern) observers are denoted by red (blue) dots and lines. Note that $13.3$ MeV/n Fe is chosen in the lower row.   The statistical significance t-test was applied and $p$-value is shown on the bottom right of each panel.   See text for details.  }
    \label{fig:observations}
 \end{figure}

Now we compare the model results of ions with measurements. The Solar Isotope Spectrometer (SIS; \citeA{Stone1998}), and the Ultra-Low Energy Isotope Spectrometer (ULEIS; \citeA{mason1998ultra}) on board the Advanced Composition Explorer (ACE) are used to survey $\sim{}$0.1 – 35\ MeV/n energetic Helium, $\sim{}$0.1 – 76\ MeV/n Oxygen, and $\sim{}$0.034 – 31\ MeV/n Iron ion observations during ESP events identified between August 1997 and December 2019. The list of events used in this study is obtained from the Near-Earth interplanetary CME (ICME) catalog (hereafter referred to as List 1), compiled by \citeA{Richardson2010}, the list of IP shocks observed during solar cycle 23 (List 2) in \citeA{Gopalswamy2010} and the GMU CME/ICME List (List 3) \cite{Regnault2020}. List 1 provides disturbance times that are typically associated with the arrival of a shock at Earth or an observing spacecraft starting from May 1996. List 2 and List 3 provide the CME originating flare position on the solar surface based on events from the first list. The total number of candidate events is 297. We then applied selection criteria to identify suitable events for our study. Specifically, we choose events where the energetic particle intensity profiles exhibited a synchronized increase of at least $200\%$ above the pre-event background within $\pm5$ hours of the shock arrival time in most of the energy bins, and no other dominant shocks were detected within a 1-day window. Using these restrictions, we identify 85 events that have the IP shock speed and SC-flare angle available. The list of events in this work is available at \url{https://doi.org/10.5281/zenodo.7969720}  \cite{ding_dataset}.

Figure~\ref{fig:observations} displays the peak intensity at $0.2$ MeV/n, $1.1$ MeV/n and $8.5$ MeV/n for He, O and Fe as a function of interplanetary (IP) shock speed. The data of $8.5$ MeV/n Fe is unavailable, we replace it with $13.3$ MeV/n. Using Equation~(\ref{eq:loglogfit}), we fit the peak intensity of the western observers ($\Delta \phi < -30^{\circ}$) and eastern observers ($\Delta \phi > 30^{\circ}$), denoted by the red and blue lines and dots respectively. Because there is an uncertainty of the CME propagation direction in observations, we choose the criteria of $\pm 30^{\circ}$ to better distinguish the eastern and western flanks.  This choice is larger than the criteria of $\pm 20^{\circ}$ in the model calculation. With the criteria of $\pm 30^{\circ}$, we have 25 events from the eastern observers, 14 from the western observers, and 46 from the central observers. We note that the number of data points in each panel is not exactly the same, since there is no significant enhancement of particle intensity for high energy channels in some events.
The central observers' peak intensities ($-30^{\circ} < \Delta \phi < 30^{\circ}$) are not displayed in this plot but can be found in \ref{appendix:central events}. Because most ESP events are observed by central observers, these events can be caused by CME eruptions that have a wide range of scales (e.g., CME width and speed) and can vary by several orders of magnitude of peak intensity at a similar speed. This large variation in peak intensity can make it difficult to accurately fit the data. However, the eastern and western ESP events are usually generated by wide shocks, which are  associated with large CME eruptions. These shocks typically have different shock geometry between the eastern and western flanks. Therefore, we limit our discussion to the possible fitting trends of the eastern and western ESP events as shown in Figure~\ref{fig:observations}. 

We conduct the two-sample t-test to examine the significance of difference between the eastern and western ESP events. In the panels of $1.1$ MeV/n, the t-test shows that at a 95\% confidence level the mean of peak intensity for eastern observers is larger than that for western observers. However, other panels show a larger $p$-value. Therefore, we can not conclusively determine the significance of difference between eastern and western data.  Nevertheless, the low $p$-value indicates a possible trend where the peak intensities may differ between eastern and western observers.    From the linear fit, we find that the fitted peak intensity of eastern observers (blue lines) is generally higher than that of western observers (red lines),  and the slope $C_{1}$ is larger at the western observers for $0.2$ MeV/n, $1.1$ MeV/n and $8.5$ MeV/n ions. These features are similar to our model calculations. As discussed above, the injection efficiency is higher at the quasi-parallel shock, which is generally associated with eastern shock flanks.  Hence, we suggest that this East-West asymmetry is a consequence of asymmetric injection efficiency.   However, the uncertainty of $C_1$ and $C_0$ is relatively large, and the coefficient of determination $R^{2}$ is small for the eastern events. Additionally, peak intensity for events having a similar shock speed shows a difference up to 2-3 magnitude. There are several possible reasons that lead to such a large variation. First, the spectral indices of the seed population may differ between events and even differ during an individual event. A recent study by \citeA{Wijsen2023JGRA..12831203W} suggests that the spectral indices of the seed population can be strongly altered by velocity shears in the solar wind (e.g., corotating interaction regions). As discussed in \ref{appendix:injection}, the injection efficiency sensitively depends on the spectral indices of the seed population.  Second, the variation of shock compression ratio and shock obliquity can also contribute to this large uncertainty as suggested by Equation~(\ref{eq:injection_efficy}). Third, the magnetic fluctuations can alter the injection energy as well as the maximum particle energy, indicated by Equation~(\ref{eq:anisotropy}),  which can in turn affect the peak intensity of ions. We note that the number of eastern and western events is limited and not enough to conduct a comprehensive statistical analysis. The observations from multiple spacecraft are necessary to increase the statistical significance of the results and to confirm the observed East-West asymmetry. However, the current study provides valuable insight into the possible mechanisms behind the observed asymmetry and can serve as a basis for future studies.

\section{Conclusion} \label{sec:conclusion} 

Recent observations \cite{ebert2016ApJ...831..153E,Duenas2022} have shown that there is an East-West asymmetry of the peak intensity in ESP events, which is an asymmetric  longitudinal distribution of the peak intensity relative to the flare center. In this work, we use the iPATH model to examine this asymmetry of the peak intensity in ESP events. Our results show that the injection efficiency peaks to the east of the CME center where the shock geometry is quasi-parallel. Consequently, for a similar shock speed, the quasi-parallel portion of a shock corresponds to a higher particle intensity than the quasi-perpendicular portion of a shock. We also examine the $Q/A$ dependence of this asymmetry for Helium, Oxygen and Iron. These heavy ions display a similar East-West asymmetry for low energies (e.g., $0.2$ MeV). However, this asymmetry of ions varies significantly at high energies because the maximum energy/nucleon of ions depends on $Q/A$. Our model results are in qualitative agreement with measurements of ESP events from the ACE observations. We note that the number of ESP events at the eastern and western shock flank is still not sufficient for a comprehensive statistical study, and we have not considered the multiple-spacecraft observations of ESP events in this work. We hope that the possible joint observations from multiple spacecraft during solar cycle 25 can provide more evidence to reveal the East-West asymmetry of ESP events.

In summary, our study suggests that injection efficiency can be the key factor leading to the East-West asymmetry of the peak intensity in ESP events, which depends sensitively on the shock obliquity. Additionally, the $Q/A$ dependence of the maximum particle energy affects this asymmetry for heavy ions.

\section{Open Research} 
ACE data are publicly available at the CDAWeb database (\url{https://cdaweb.gsfc.nasa.gov/index.html/}). All simulation data presented in the paper and the list of ESP events are available at Zenodo via \url{https://doi.org/10.5281/zenodo.7969720} \cite{ding_dataset}.

\acknowledgments
This work is supported in part by NASA grants 80NSSC19K0075, 80NSSC19K0831 and 80NSSC19K0079 at UAH. Work at SwRI is supported by NASA grants 80NSSC20K1815 and NNX17AI17G. Supports by ISSI and ISSI-BJ through the international teams 469 is also acknowledged. N.W.\ acknowledges support from the NASA program NNH17ZDA001N-LWS and from the Research Foundation -- Flanders (FWO -- Vlaanderen, fellowship no.\ 1184319N). The work at KU Leuven has received funding from the European Union’s Horizon 2020 research and innovation programme under grant agreement No 870405 (EUHFORIA 2.0) and the ESA project “Heliospheric modelling techniques" (Contract No. 4000133080/20/NL/CRS). These results were also obtained in the framework of the projects C14/19/089  (C1 project Internal Funds KU Leuven), G.0D07.19N  (FWO-Vlaanderen), SIDC Data Exploitation (ESA Prodex-12), and Belspo project B2/191/P1/SWiM. For the computations we used the infrastructure of the VSC-Flemish Supercomputer Center, funded by the Hercules foundation and the Flemish Government-department EWI.

\appendix

\section{Injection energy and injection efficiency} \label{appendix:injection}

\begin{figure}[ht]
    \centering
    \includegraphics[width=\textwidth]{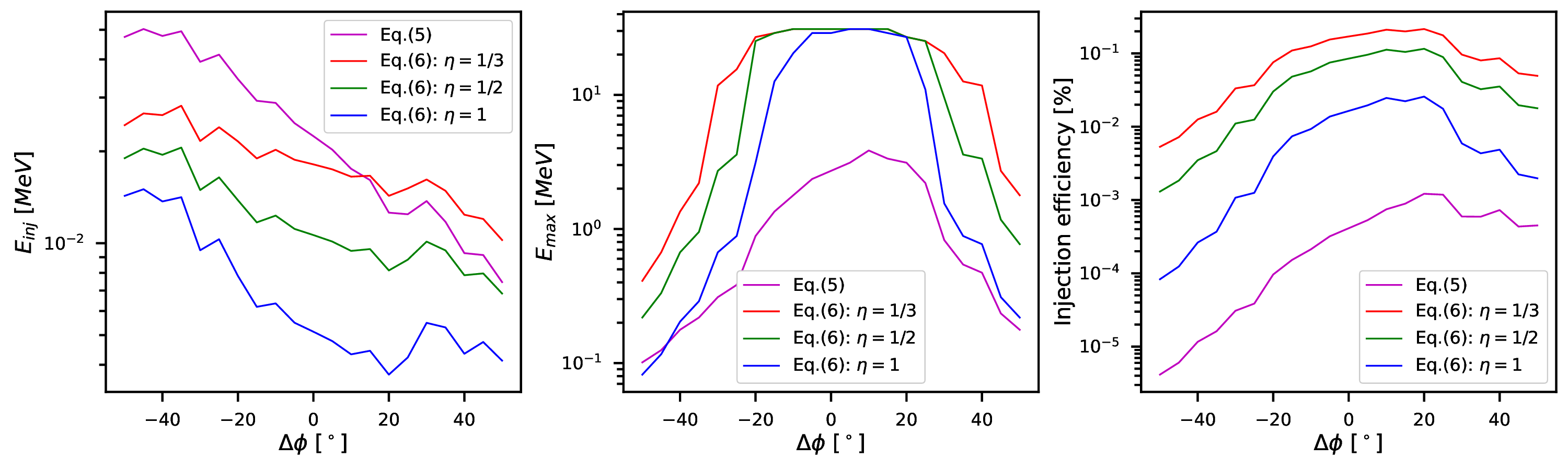}
    \caption{The injection proton energy, the maximum proton energy and injection efficiency as a function of SC-CME deflection angle ($\Delta \phi$) at 1 au. The labels correspond to the equations of injection speed in Section~\ref{sec:model}. See text for details.  }
    \label{fig:appdix_injection}
 \end{figure}

 By way of example, we use the shock parameters for the case shown in Figure~\ref{fig:cme-shock} to compare the different formulae for injection speed and the associated maximum proton energy at $1$ au. As discussed in Section~\ref{sec:model}, the injection efficiency depends on the choice of injection formulae and so is the  maximum particle energy since the amplified wave strength is proportional to the number of injected particles. We calculate injection energy, the maximum proton energy and the injection efficiency based on Equation~(\ref{eq:injection speed_zank}) and Equation~(\ref{eq:injection speed_eta}).  We consider  a strong parallel shock, with $s=4$ and $\theta_{\rm BN}=0^{\circ}$, as the base case.  Its  injection efficiency is set to $\epsilon_{0} =1\%$ and we use $\delta=3.5$. Then the injection efficiency at oblique shocks can be obtained accordingly. The injection efficiency $\epsilon_{1}$  associated with Equation~(\ref{eq:injection speed_zank}) at oblique shocks is given by,
 \begin{equation}
 \epsilon_{1}^{Eq. 5} = \epsilon_{0} \left(\beta-3 + 3\tan\theta_{\rm BN}\right)^{2-2\delta}.
  \label{eq:injection_efficy_zank}
 \end{equation} 
 Similarly, the injection efficiency $\epsilon_{1}$ from Equation~(\ref{eq:injection speed_eta}) can be expressed as 
 \begin{equation}
 \epsilon_{1}^{Eq. 6}  = \epsilon_{0} \left(\frac{(\beta-3\eta)+\tan\theta_{\rm BN}}{4-3\eta}\right)^{2-2\delta}.
  \label{eq:injection_efficy_eta}
 \end{equation}
 Figure~\ref{fig:appdix_injection} compares the injection energy, the maximum proton energy and the injection efficiency as a function of SC-CME deflection angle ($\Delta \phi$) using Equation~(\ref{eq:injection speed_zank}) and Equation~(\ref{eq:injection speed_eta}). For Equation~(\ref{eq:injection speed_eta}), $\eta=1$, $\eta=1/2$ and $\eta=1/3$ are chosen. Lower values of $\eta$ result in higher injection energy ($E_{\rm inj}$) and maximum energy ($E_{\rm max}$) due to their strong influence on the injection efficiency, as demonstrated in the right panel. Equation~(\ref{eq:injection speed_zank}) has the lowest injection efficiency since its injection energy has a stronger dependence on shock obliquity angle. For a reasonable estimate of the injection efficiency and its effects on the amplified wave intensity, we adopt Equation~({\ref{eq:injection speed_eta}}) with $\eta=1/3$ as the injection speed threshold in the iPATH model.

\begin{figure}[ht]
    \centering
    \includegraphics[width=\textwidth]{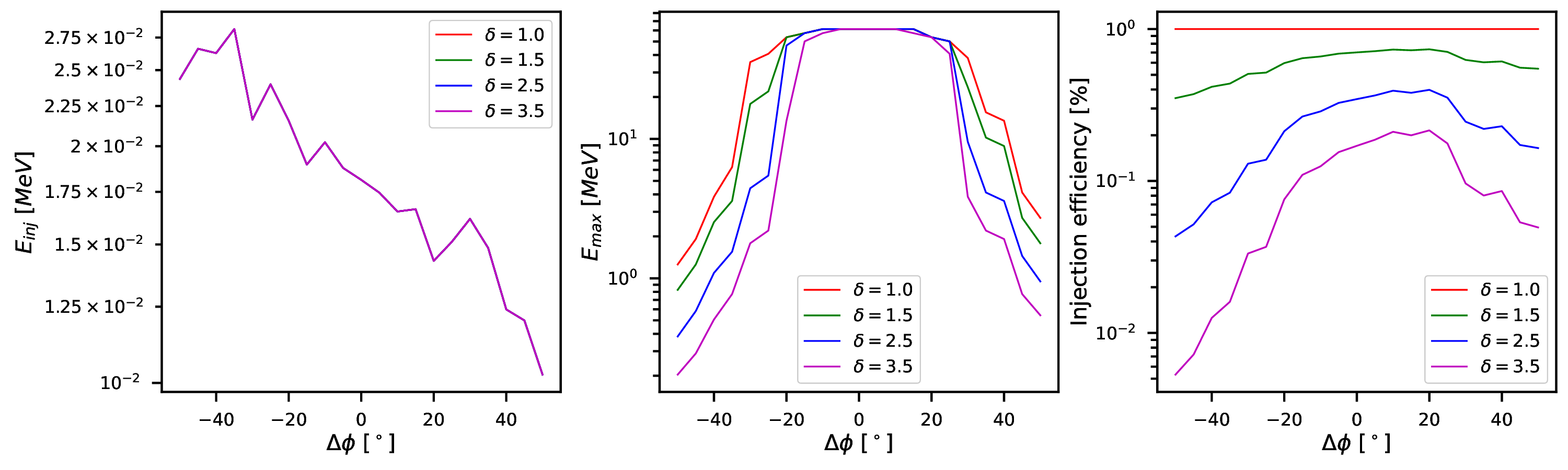}
    \caption{The injection proton energy, the maximum proton energy and injection efficiency as a function of SC-CME deflection angle ($\Delta \phi$) at 1 au. The labels represent the spectral index $\delta$ of the seed population. }
    \label{fig:appdix_injection_seed_spec}
    
 \end{figure}

 In addition to the shock properties, the injection efficiency and its East-West asymmetry are also influenced by the spectral index of the seed population. Figure~\ref{fig:appdix_injection_seed_spec} illustrates the comparison of the maximum proton energy and injection efficiency as a function of $\Delta \phi$ for different spectral indices ($\delta=1,1.5,2.5,3.5$). The injection energy is the same for all cases using Equation~(\ref{eq:injection speed_eta}) with $\eta=1/3$. The injection efficiency decreases with an increasing spectral index, leading to a decrease in the maximum proton energy at shock flanks. The plateau of $E_{\rm max}$ near the shock nose is due to the wave intensity reaching the Bohm limit \cite{Zank+etal+2000}. Note that the injection efficiency exhibits a more significant East-West asymmetry for larger $\delta$, while there is no asymmetry of injection efficiency when $\delta=1$.

\section{Supplementary Figure on ESP intensity}\label{appendix:central events}

\begin{figure}[ht]
    \centering
    \includegraphics[width=\textwidth]{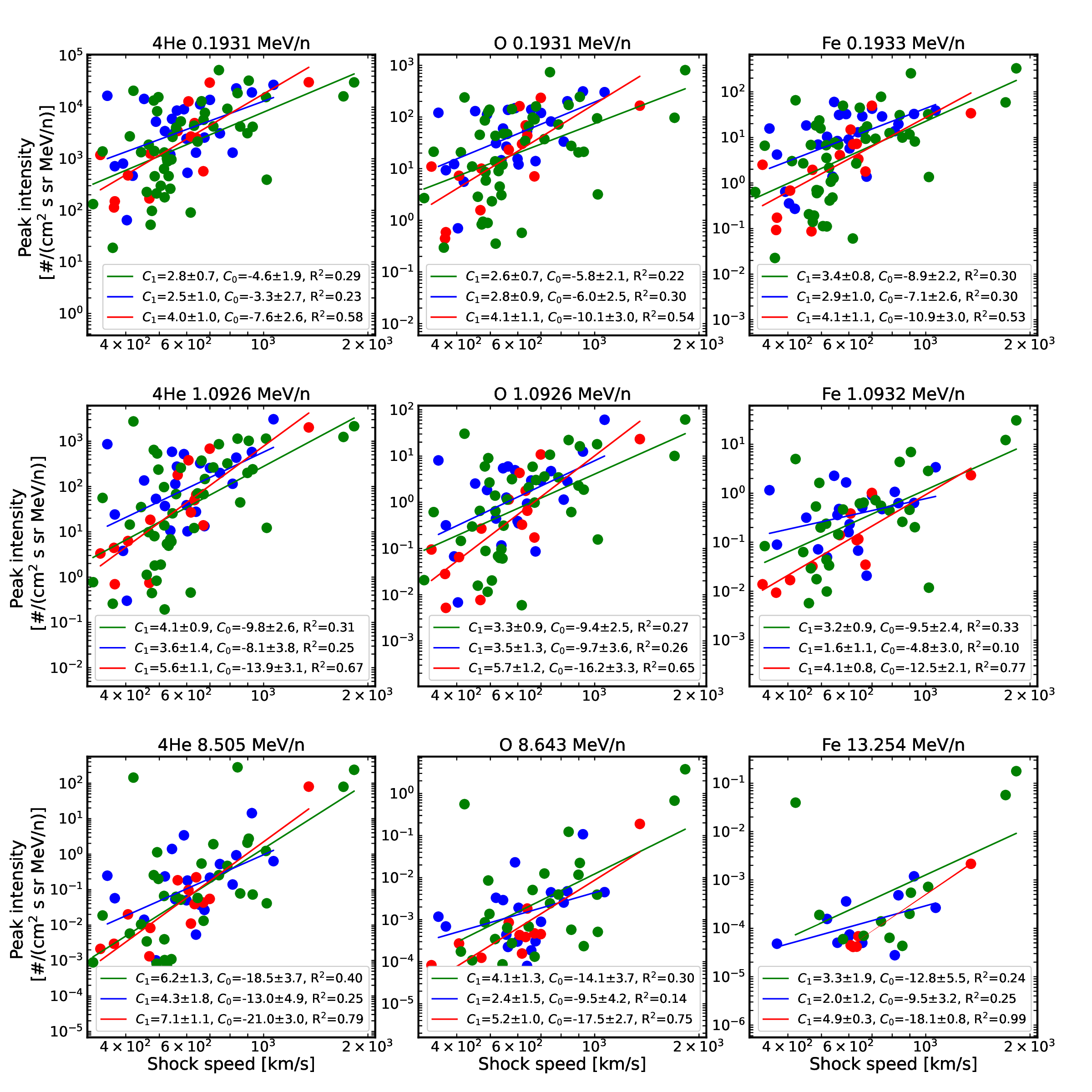}
    \caption{Peak intensity versus the IP shock speed at $0.2$ MeV/n (upper row), $1.1$ MeV/n (middle row) and $8.5$ MeV/n (lower row) for He, O and Fe in the observations.
    (except the lower right panel, where $13.3$ MeV/n is used for Fe). Western, eastern and central observers satisfy
    $\Delta \phi < -30^{\circ}$, $\Delta \phi > 30^{\circ}$, and
    $-30^{\circ} < \Delta \phi < 30^{\circ}$,
    and are denoted by red, blue and green dots and lines, respectively.  }
    \label{fig:observations-appdx}
 \end{figure}

Figure~\ref{fig:observations-appdx} provides additional information on the peak intensity of ESP events at western, eastern and central observers, denoted by red, blue and green dots and lines respectively. The central events, shown as the green dots, with $-30^{\circ} < \Delta \phi < 30^{\circ}$, have the largest number of data points. These events are often associated with the passage of the shock nose, which has the highest likelihood of generating ESP events. However, the peak intensities can vary by several orders of magnitude at a similar IP shock speed. As shown in Figure~\ref{fig:observations-appdx}, most ESP events are observed by central observers and some small ESP events can also be recorded by central observers (i.e., events with low peak intensity). The large variability of peak intensity may indicate the complexity and diversity of shock parameters for central events. In comparison, eastern and western events are usually caused by large SEP events where wide CME-driven shocks are often observed. The variability of these events are therefore smaller than the central events.  
For instance, the sample of eastern and western events of $0.2$ MeV/n ions (blue and red dots) tends to have a higher peak intensity, whereas there are many central events with a low peak intensity.
Because the central events may include many smaller events, we exclude them in this study and focus on the western and eastern events which typically have distinct differences in shock obliquity.

%
%


\bibliography{esp}

%
%
%
%
%

\end{document}